\documentstyle[11pt,aaspp4,flushrt,psfig]{article}

\begin{document}

\title{Numerical stability of a family of Osipkov-Merritt models}
\author{Andr\'es Meza and Nelson Zamorano}
\affil{Universidad de Chile, Facultad de Ciencias F\'{\i}sicas y
Matem\'aticas, \\ Departamento de F\'{\i}sica, Casilla 487--3, Santiago,
Chile}

\begin{abstract}
We have investigated the stability of a set of non-rotating anisotropic
spherical models with a phase-space distribution function of the
Osipkov-Merritt type. The velocity distribution in these models is isotropic
near the center and becomes radially anisotropic at large radii. They are
special members of the family studied by Dehnen and Tremaine et al. where
the mass density has a power-law cusp $\rho\propto r^{-\gamma}$ at small
radii and decays as $\rho\propto r^{-4}$ at large radii.

The radial-orbit instability of models with $\gamma$ = 0, 1/2, 1, 3/2, and
2, was studied using an N-body code written by one of us and based on the
`self-consistent field' method developed by Hernquist and Ostriker. These
simulations have allowed us to delineate a boundary in the
$(\gamma,r_{a})$-plane that separates the stable from the unstable models. 
This boundary is given by $2T_{r}/T_{t} = 2.31 \pm 0.27$, for the ratio of
the total radial to tangential kinetic energy. We also found that the
stability criterion $df/dQ\le 0$, recently raised by Hjorth, gives lower
values compared with our numerical results.

The stability to radial modes of some Osipkov-Merritt $\gamma$-models which
fail to satisfy the Doremus-Feix's criterion, $\partial f/\partial E<0$, has
been studied using the same N-body code, but retaining only the $l=0$ terms
in the potential expansion. We have found no signs of radial instabilities
for these models.

\end{abstract}


\newpage 

\section{Introduction}

Spherical stellar systems in which the orbits are strongly eccentric, or
radial, are unstable to forming a bar. This so-called radial-orbit
instability (hereafter ROI), was first demonstrated for a sphere consisting
entirely of radial orbits by Antonov (1973). It was actually observed by
H\'enon (1973) in a series of N-body simulations for the generalized
polytropes, defined by a distribution function which is the product of power
laws in energy and angular momentum. Subsequent N-body simulations (Merritt
\& Aguilar 1985; Barnes, Goodman, \& Hut 1986; Dejonghe \& Merritt 1988) and
numerical linear stability analysis (Saha 1991; Weinberg 1991) have revealed
that a small velocity anisotropy can be enough to make a system unstable;
they also have showed that such a system evolves quickly into a triaxial
bar.

The ROI can be explained by the transformation of loop orbits into box
orbits when the spherical symmetry of the potential is broken (Merritt
1987). The mechanism is similar to the scenario described by Lynden-Bell
(1979) for the formation of a bar in a disk. Initially, all the orbits are
precessing ellipses (loops). A weak bar-like ($l = 2$) perturbation
generates a small torque that could trap those orbits with the lowest
angular momenta; realign them around the bar, reinforcing the strength of
the initial perturbation. This instability implies an upper limit to the
degree of velocity anisotropy in spherical models. Dejonghe \& Merritt
(1988) studied the stability of the Plummer (1911) model with two different
anisotropic distribution function (hereafter DF) and concluded that the ROI
is the most robust and is likely to be the most important instability in
elliptical galaxies.

For spherical stellar systems with isotropic DF, $f=f(E)$, it is known that
$df/dE<0$ acts as a sufficient condition for the stability under radial and
nonradial perturbations (Antonov 1962; Sygnet et al. 1984). However, the
situation for anisotropic spherical models is poorly understood and only the
stability to radial modes can be analytically tested using the sufficient
condition $\partial f/\partial E<0$ (Dor\'emus \& Feix 1973). A sufficient
stability criterion based in the global anisotropy parameter $2T_{r}/T_{t}$,
the ratio of kinetic energies corresponding to the radial and tangential
direction, has been suggested by Polyachenko \& Shukhman (1981; see Fridman
\& Polyachenko 1984). However, N-body simulations have showed that is not a
reliable criterion because its value seems to be model dependent (Merritt \&
Aguilar 1985; Dejonghe \& Merritt 1988). Moreover, Palmer \& Papaloizou
(1987) have analytically demonstrated that no stability criterion can be
found using the anisotropy parameter $2T_{r}/T_{t}$, because there is always
a spectrum of unstable (growing) modes if the DF becomes singular for a
vanishing angular momentum; no matter how weak appears the DF divergence.

Recently, Hjorth (1994) has proposed a simple criterion for the stability of
models with DF of the Osipkov-Merritt type (Osipkov 1979; Merritt 1985).
These models are described by an anisotropic DF, $f(E,L^{2}) = f(Q)$, with
$Q \equiv E - L^{2}/2r_{a}^{2}$; where $E$ is the energy per unit mass and
$L$ is the magnitude of the angular momentum per unit mass. Inside the
anisotropy radius $r_{a}$ the velocity distribution is nearly isotropic,
while outside $r_{a}$ the radial anisotropy increases. Thus, the parameter
$r_{a}$ controls the global anisotropy degree in the velocity distribution;
the smaller its value, the larger the anisotropy of the system.

In this paper we show the results of a numerical study of the stability
properties for a set of Osipkov-Merritt models. These are special members of
the family of spherical $\gamma$-models independently studied by Dehnen
(1993, hereafter D93) and Tremaine et al. (1994, hereafter T94); with
densities that behave as $\rho\propto r^{-\gamma}$ near the center and as
$\rho\propto r^{-4}$ for large radii. The stability was investigated using
an N-body code based on the `self-consistent field' method developed by
Hernquist \& Ostriker (1992, hereafter HO92).

In \S\ 2 we briefly describe the $\gamma$-models and the N-body code used to
study their evolution. In \S\ 3 we present our results for models with
$\gamma$ = 0, 1/2, 1, 3/2, and 2. In particular, we are able to establish a
stability boundary for the ROI in the $(\gamma,r_{a})$-plane. A discussion and 
summary of our results appear in \S\ 4.

\section{N-body Simulations}

\subsection{Models and Initial Conditions \label{sec:models}}

In this section, we briefly summarize the properties of the one-parameter
family of spherical models studied by D93 and T94 (more details can be found
in the references just quoted). These $\gamma$-models have a mass density
given by
\begin{equation}
\rho(r) = \frac{3-\gamma}{4\pi} \frac{aM}{r^\gamma(r+a)^{4-\gamma}}, \;\;\; 0
\le \gamma < 3,
\label{eq:density}
\end{equation}
where $a$ is a typical scale length and $M$ is the total mass of the system
(T94 describe the same density by the parameter $\eta = 3 - \gamma$ and call
them ``$\eta$-models''). The models of Jaffe (1983) and Hernquist (1990,
hereafter H90) are recovered for $\gamma = 2$ and $\gamma = 1$, respectively.

The gravitational potential associated with the density (\ref{eq:density}) is
obtained from the Poisson equation, and has the simple form
\begin{equation}
\Phi(r) = \frac{GM}{a} \times \left\{ \begin{array}{ll}
     \displaystyle
     \ln\left(\frac{r}{r+a}\right), & \mbox{for $\gamma=2$,} \\
     & \\
     \displaystyle
    -\frac{1}{2-\gamma}\left[1-\left(\frac{r}{r+a}\right)^{2-\gamma}\right],
     & \mbox{for $\gamma\neq 2$},
    \end{array} \right.
\label{eq:potential}
\end{equation}
where $G$ is the gravitational constant. In the following, we adopt a system
of units in which $M$, $a$, and $G$ are unity.

Many dynamical properties of the $\gamma$-models with an isotropic DF, $f =
f(E)$, were given by D93 and T94. The former also gave analytical
expressions for anisotropic DFs of the Osipkov-Merritt type for some
$\gamma$-models.  More recently, Carollo, de Zeeuw, \& van der Marel (1995)
have been able to express the projected line-of-sight velocity profiles of
models with $\gamma \le 2$ in a single quadrature. Models with $\gamma > 2$
are less suited to describe spherical galaxies because they display some
unrealistic properties, such as infinite central potential and surface
density profiles that differ significantly from the $R^{1/4}$ law (see D93).

We have limited our study to models with $\gamma$ = 0, 1/2, 1, 3/2, and 2,
which represent a wide range for the density cusp at small radii. The
$\gamma = 0$ model is the only one of this family that has a
(non-isothermal) core.  The model that most closely resembles the $R^{1/4}$
law in surface density is the $\gamma = 3/2$ model (D93). In spite of a
previous study by Merritt \& Aguilar (1985) of Jaffe ($\gamma = 2$) model,
we have included it here to obtain a more accurate estimation for its
stability threshold. In fact, as we show below, we found a slightly larger
value for the critical anisotropy radius than theirs.

For all these models, we have generated a set of initial conditions using
the Osipkov-Merritt DF\footnote{There is a factor 2 missing in the second
term of the right-hand side of equation (A7) of D93.}. Each model was
truncated at a radius enclosing 99.9\% of the total mass. All simulations
employ $N=50,000$ equal mass particles. Table \ref{table1} shows the
parameters associated with each of these models; where $r_{h}$ is the
half-mass radius, $r_{0}$ is the lower value for the anisotropy radius such
that the DF is strictly positive, and $T_{h}$ is the dynamical time
evaluated at $r_{h}$.
 
\subsection{N-body code \label{sec:code}}

The stability of these models has been studied using an N-body code based on
the `self-consistent field' method developed by HO92 (see also Clutton-Brock
1973). In this approach, the density and the gravitational potential are
expanded in a biorthogonal set of basis functions as
\begin{equation}
\begin{array}{lclcl}
\rho({\bf r}) & = & \displaystyle \sum_{nlm} A_{nlm}\rho_{nlm}({\bf r}) & = &
\displaystyle \sum_{nlm} A_{nlm}\tilde{\rho}_{nl}(r)Y_{lm}(\theta,\varphi), \\
&&&& \\
\Phi({\bf r}) & = & \displaystyle \sum_{nlm} A_{nlm}\Phi_{nlm}({\bf r}) & = &
\displaystyle \sum_{nlm} A_{nlm}\tilde{\Phi}_{nl}(r)Y_{lm}(\theta,\varphi).
\end{array}
\label{eq:expansion}
\end{equation}

HO92 obtained their basis functions $\{\rho_{nlm},\Phi_{nlm}\}$ using the
model for spherical galaxies proposed by H90 as the zeroth-order term. Higher
order terms are found by construction. For a set of $N$ pointlike particles
the expansion coefficients $A_{nlm}$ are given in terms of the positions of
the particles. Then, the accelerations are computed by a simple analytical
differentiation of the potential expansion (see HO92 for a more detailed
description).

One of us (AM) has written a code that  implements this formalism. In this
code the particle positions and velocities are updated using a second order
integrator with a fixed timestep $\Delta t$, given by
\begin{eqnarray}
{\bf x}_{i+1} & = & {\bf x}_{i} + \Delta t \, {\bf v}_{i} +
\frac{1}{2} \,  \Delta t^{2} \, {\bf a}_{i}, \\
&&\nonumber \\
{\bf v}_{i+1} & = & {\bf v}_{i} + \frac{1}{2} \, \Delta t \,
({\bf a}_{i} + {\bf a}_{i+1}),
\end{eqnarray}
where the subscript identifies the iteration. This integrator is equivalent 
to the standard time centered leapfrog, as can be verified by direct 
substitution (Allen \& Tildesley 1992; Hut, Makino, \& McMillan 1995).
We have studied its numerical properties and found them closely similar to
those of the standard leapfrog.

The HO92 basis set can be used to represent a number of potential-density
pairs for spherical galaxies, including the $\gamma$-models. For example,
the $\gamma = 1$ model is just the zeroth order term and the $\gamma = 0$
model can be reproduced exactly as a linear combination of two terms of this
basis (HO92). However, for practical purposes, the potential (density) of
other $\gamma$-models are approximated using truncated expansions of the
kind defined in equation (\ref{eq:expansion}) limiting the number of radial
functions to $n_{\rm max}$.

\begin{figure}[t]
\plotone{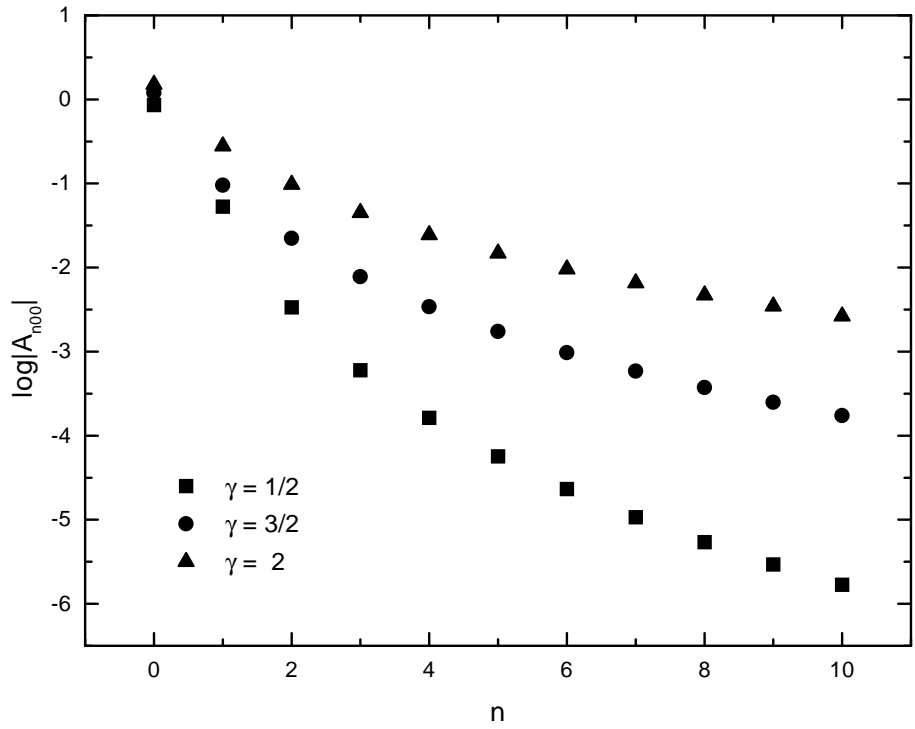}
\caption{Expansion coefficients for some $\gamma$-models using
the HO92 basis set. \label{fig:coeff}}
\end{figure}

\begin{figure}[t]
\centerline{\psfig{figure=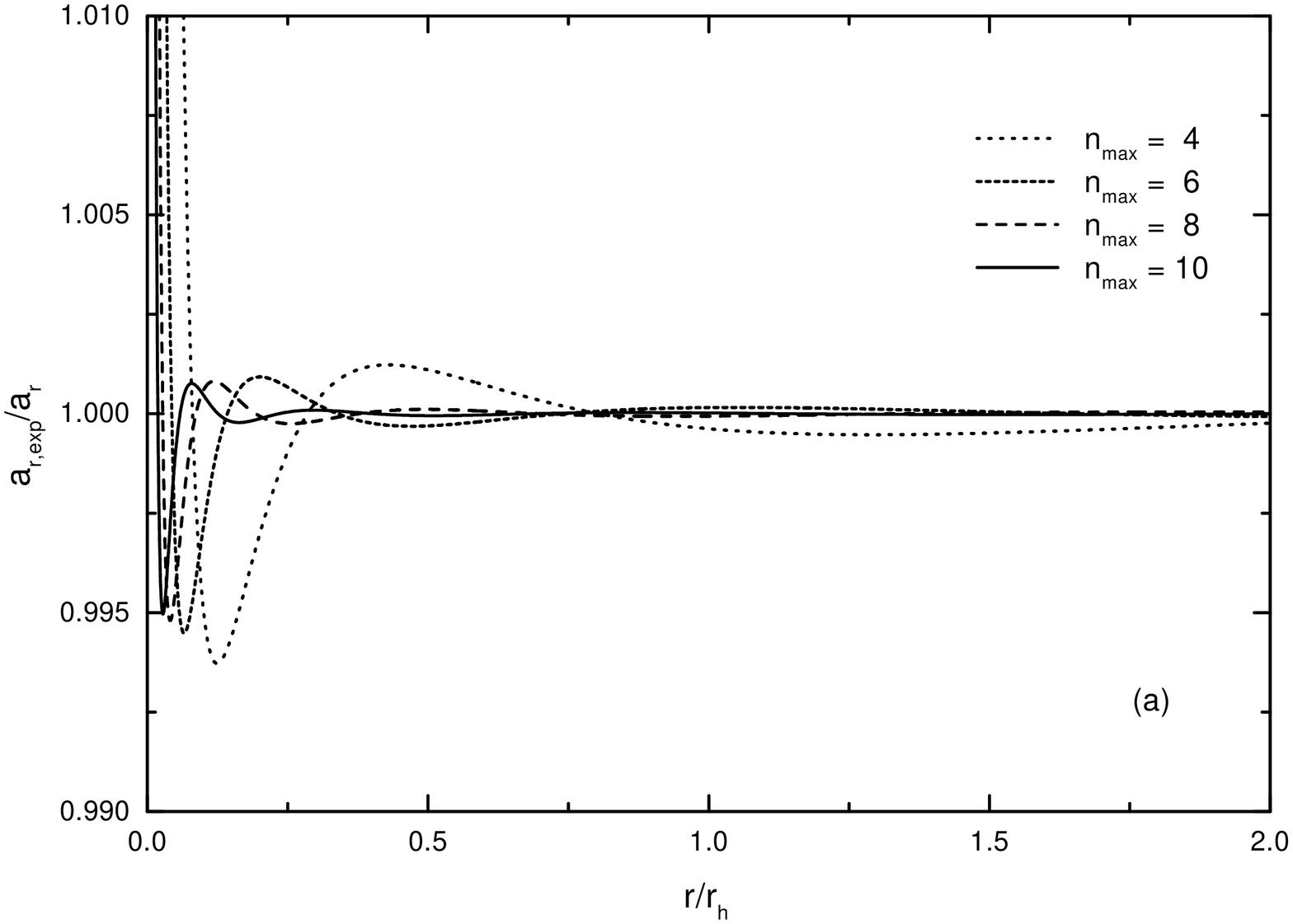,height=0.55\hsize,angle=0}}
\centerline{\psfig{figure=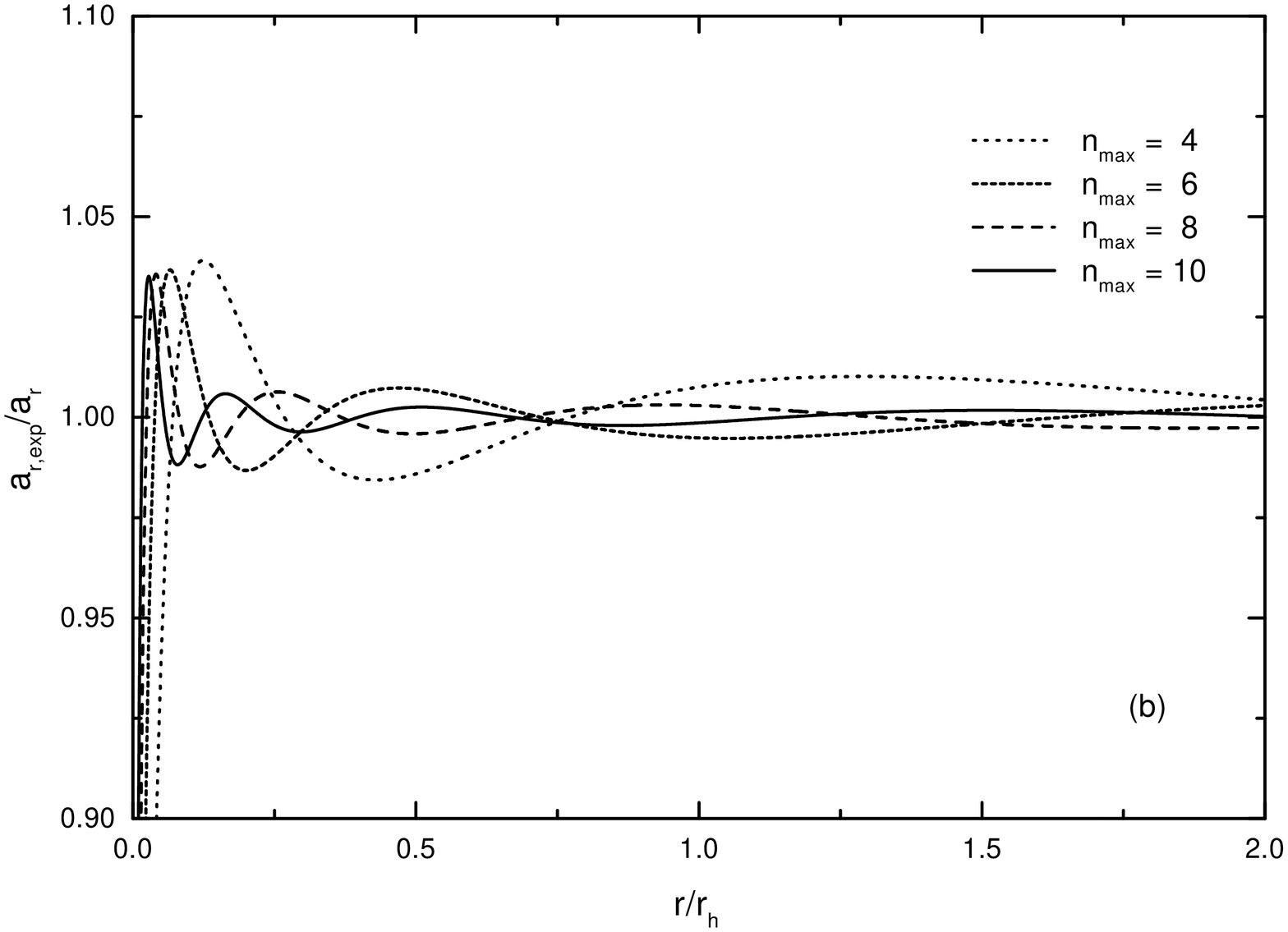,height=0.55\hsize,angle=0}}
\caption{Relative error in the radial acceleration for the
$\gamma$ = 1/2 (a) and $\gamma$ = 3/2 (b) model as a function of radius $r$.
The figures compare the acceleration, as computed using the HO92 basis
functions for different values of $n_{\rm max}$, to their exact values.
\label{fig:accel}}
\end{figure}

In our simulations, $n_{\rm max}$ has been chosen according to the studied
model. Figure \ref{fig:coeff} shows the values of the expansion coefficients
$A_{n00}$ (see eq. [2.35] of HO92) for some of the $\gamma$-models. The 
coefficients are directly related with $\gamma$; smaller values of $A_{n00}$ 
are obtained for lower $\gamma$. In all the cases, for large values of $n$ 
the coefficients $A_{n00}$ levels off. Then, adding more terms to the 
potential expansion does not decrease significantly the error, but the 
computational time increases considerably. However, in most cases, good 
accuracy in the evaluation of the potential can be achieved with 
$n_{\rm max} \sim 6$. This point is illustrated in Figure \ref{fig:accel}, 
where we compare the radial acceleration computed using the HO92 basis
functions for the potential expansion (\ref{eq:expansion}), truncated at
different values of $n_{\rm max}$, to its exact value for the $\gamma$ = 1/2
and $\gamma$ = 3/2 models. In both cases, at large radii, the acceleration
obtained using the truncated potential expansion agrees with the one
obtained using the full potential. However, at small radii, the error grows
with $\gamma$. For $\gamma$ = 1/2, the radial acceleration can be computed
to better than 0.5\% accuracy at small radii, while for $\gamma$ = 3/2 the
same quantity can be evaluated only up to 5\% accuracy. The adopted values
for $n_{\rm max}$ (see Table \ref{table1}) were chosen such that the error
in the acceleration, at small radii, was lower that 5\%.

Since we are interested in the ROI, all our simulations only include series
expansions up to $l_{\rm max} = 2$. The total elapsed time was chosen as
50 half-mass dynamical times. The timestep for the different models appears
in Table \ref{table1}. They were taken such that the energy was
conserved better than 0.5\% during the total elapsed time. In some special
cases, we checked that our results would not depend on the number of
terms used in the potential expansion and the timestep (details are given
below).

\section{Results\label{sec:results}}

To study the ROI in the models introduced in \S\ \ref{sec:models}, we have
done a set of simulations with different values for the anisotropy radius
$r_{a}$. For each run, we have used an iterative algorithm to estimate the
axial ratios of the particle distribution. In this scheme, initial values for
the modified inertia tensor
\begin{equation}
I_{ij} = \sum \frac{x_{i} x_{j}}{a^{2}}, \hspace{2cm}
a^{2} = x^{2} + y^{2}/q_{1}^{2} + z^{2}/q_{2}^{2}
\end{equation}
are calculated for all particles inside a spherical shell ($q_{1}=q_{2}=1$)
with a given radius $r_{\rm m}$. New axis ratios $q_{1}$, $q_{2}$ and the
orientation of the fitting ellipsoid are then estimated from the eigenvalues
of $I_{ij}$ and used to obtain an improved approximation to the modified
inertia tensor. This process is repeated until the axis ratios converge to a
value within a pre-established tolerance, $\Delta q=0.001$.

\begin{figure}[t]
\plotone{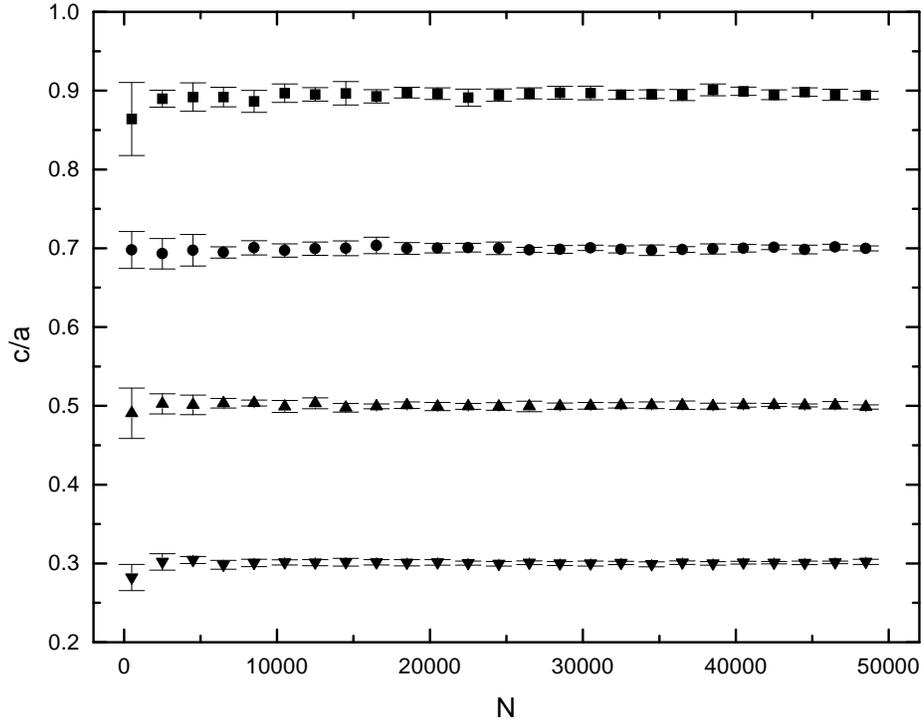}
\caption{Computed minor axis ratio as a function of the number of
particles $N$ for a set of 10 random generations of the ``triaxial'' Hernquist
profile (eq. [\ref{eq:triaxial}]) with $c = 0.3$ (down triangles), $c = 0.5$
(up triangles), $c = 0.7$ (circles), and $c = 0.9$ (squares). The error bar
quoted here corresponds to the standard deviation of the measures.
\label{fig:error} }
\end{figure}

We have tested the precision of this scheme computing the axis ratio of a set
of random $N$ particle positions generated from the following triaxial
generalization of the Hernquist model (e.g., Merritt \& Fridman 1996)
\begin{equation}
\rho(m) = \frac{1}{2\pi bc} \frac{1}{m(1+m)^3},
\label{eq:triaxial}
\end{equation}
with 
\begin{equation}
m^2 = x^2 + \frac{y^2}{b^2} + \frac{z^2}{c^2}, \hspace{2cm} 0 \le c \le b \le 1.
\end{equation}
Figure \ref{fig:error} shows the average computed minor axis ratio as a
function of the number of particles $N$ within the radius $r_{\rm m}$. The
error bar quoted correspond to the standard deviation for 10 different
measures. We found that for $N \gtrsim 1000$, the computed axis ratio of the
particle distribution came out with an error smaller than 1\%. Convergence
problems appear only when the number of particles inside the measuring
radius $r_{\rm m}$ is small, or when $r_{\rm m}$ itself is small. To avoid
this situation we have adopted as a measuring radius $r_{\rm m}$ the one
that encloses the 70\% of the total mass of the model. With these values for
$r_{\rm m}$, we have obtained the axis ratio used as a test for stability.

\begin{figure}[t]
\plotone{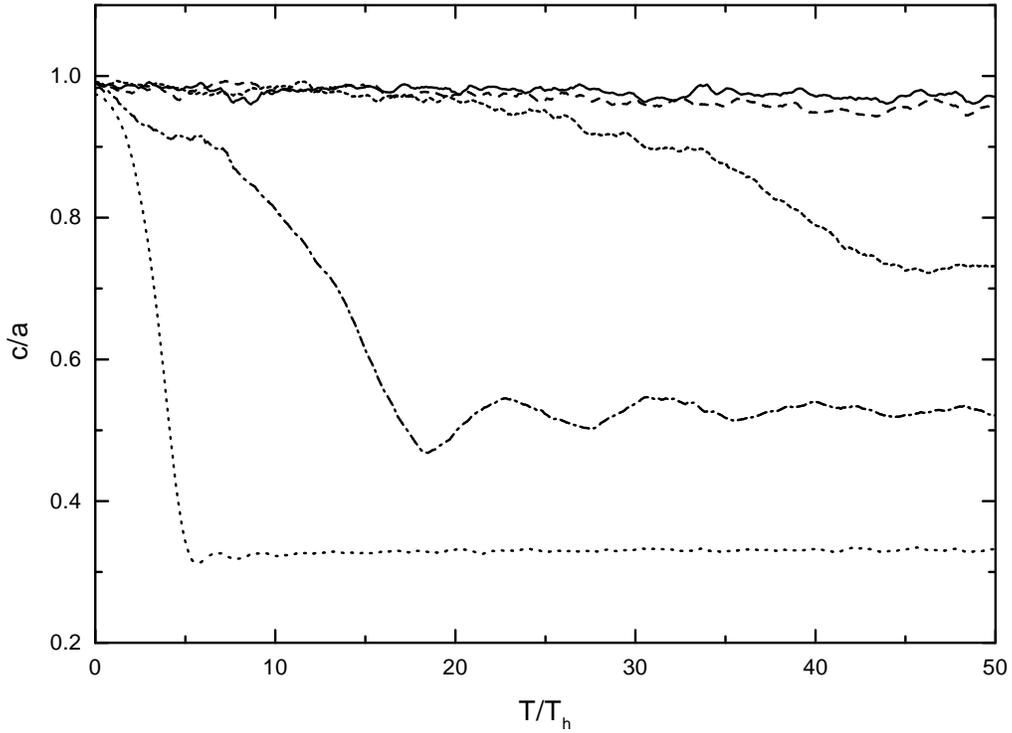}
\caption{Evolution of the minor to major axis ratio for the
Hernquist ($\gamma=1$) model for different values of the anisotropy radius
$r_{a}$. The axis ratios were obtained by fitting an ellipsoid at radius
$r_{\rm m}=5$, the radius which encloses $\sim$ 70\% of the total mass. The
time is normalized to the half-mass dynamical time $T_{h}$. Solid
line: $r_{a}=1.1$; long-dashed line: $r_{a}=1.0$; short-dashed line:
$r_{a}=0.9$; dotted-dashed line: $r_{a}=0.7$; and dotted line: $r_{a}=0.3$.
\label{fig:hernquist}}
\end{figure}

The evolution of the ratio between the minor to major axis for the Hernquist
model with different values of the anisotropy radius $r_{a}$ appears on
Figure \ref{fig:hernquist}. These values were obtained by fitting an
ellipsoid at radius $r_{\rm m} = 5$. The stability boundary appears to be
close to $r_{a} = 1.0$ and we assume that this model is stable. Models with
$r_{a}\lesssim 0.8$ are strongly unstable and their final configuration is
nearly prolate, with the final axis ratio determined by their initial
anisotropy; the larger the value of the initial anisotropy, the smaller the
final axis ratios. An example of the final state of a strongly unstable
Hernquist model appears on Figure \ref{fig:projection}, where we have
plotted the particle positions, projected along the minor semiaxis
($z$-axis), at the start and the end of a simulation for a model with
anisotropy radius $r_{a}=0.3$. The particle positions are rotated in such a
way that the semiaxis become aligned with the cartesian axis. At the end of
the run, it is clearly visible a (triaxial) bar at the center of the system
with $c/a=0.33$ and $b/a=0.40$.

\begin{figure}[t]
\plotone{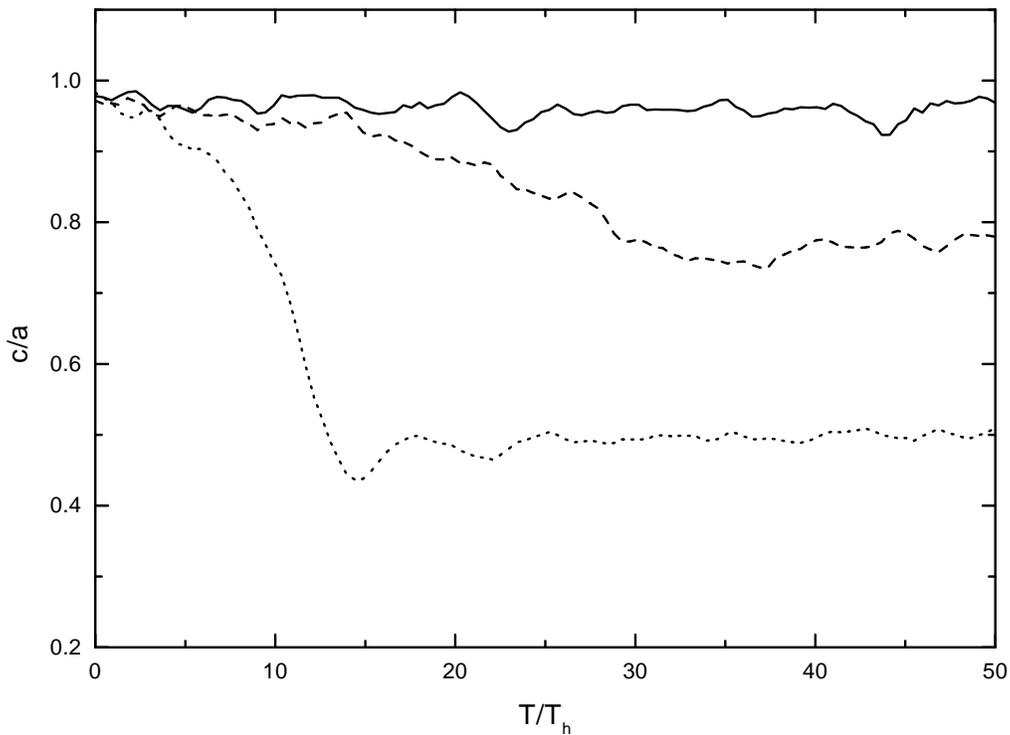}
\caption{Evolution of the minor to major axis ratio for the
Jaffe ($\gamma=2$) model. These values were obtained by fitting an ellipsoid
at radius $r_{\rm m}=2.3$, the radius which encloses $\sim$ 70\% of the
total mass. The time is normalized to the half-mass dynamical time
$T_{h}$. Solid line: $r_{a}=0.4$; dashed line: $r_{a}=0.3$; and dotted line:
$r_{a}=0.2$. \label{fig:jaffe}}
\end{figure}

Figure \ref{fig:jaffe} shows the evolution of the minor to major axis ratio
for the Jaffe ($\gamma=2$) model. The axis ratios were evaluated at radius
$r_{\rm m} = 2.3$. In their study of this model, Merritt \& Aguilar (1985)
found a critical value for $r_{a}$ between 0.2 and 0.3. However, the total
elapsed time in their simulations was only 20 dynamical times. This ambiguity
is settled in Figure \ref{fig:jaffe}, where we can see that the model with
$r_{a}=0.3$ appears to be stable for elapsed times $t \lesssim 20T_{h}$,
however, for $t \sim 30T_{h}$ the system show signs of an incipient bar.
Accordingly, we set the stability threshold at $r_{a}\sim 0.4$.

The evolution of the axis ratio for other $\gamma$-models is very similar.
Our results are summarized in Figure \ref{fig:plane}. Table \ref{table1}
includes The critical anisotropy radius $r_{a}$ and the respective critical
global anisotropy $2T_{r}/T_{t}$. The stability boundary between stable and
unstable models is given roughly by the linear fit $r_{a}=1.54-0.54\gamma$.
This boundary came out very close with the conservative estimate
$r_{a}=1.6-0.67\gamma$, raised by Carollo, de Zeeuw, \& van der Marel (1995)
based on the supposition that the stability boundary for models with 
$\gamma<2$ is given by $2T_{r}/T_{t}=2.5$, the critical global anisotropy 
quoted by Merritt \& Aguilar (1985) for the $\gamma=2$ model. 

Our results give a critical global anisotropy parameter
$2T_{r}/T_{\perp}=2.31 \pm 0.27$ for the $\gamma$-models; however, the values
of the anisotropy parameter are larger for lower values of $\gamma$. This
limit is larger than the values 1.75 and 1.58 reported by Fridman \&
Polyachenko (1984) and Bertin et al. (1994), respectively. In our opinion,
these results strengthen the belief that this criterion does not yield a
reliable prediction for the stability threshold value because it is strongly
model dependent.

\begin{figure}[t]
\plotone{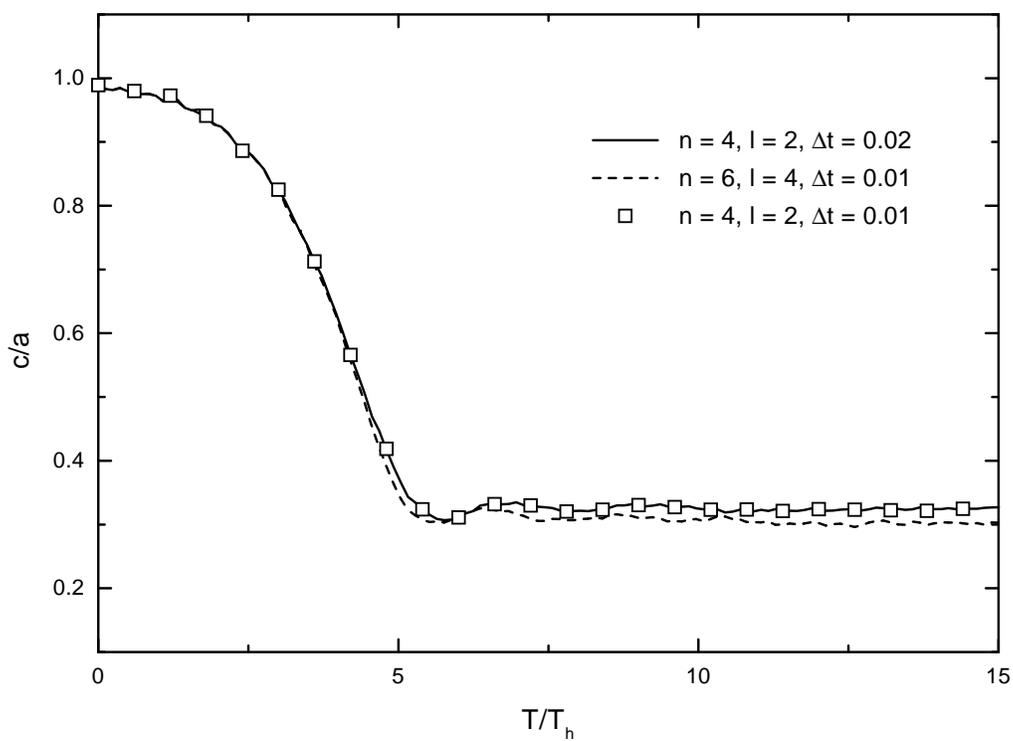}
\caption{Evolution of the minor to major axis ratio for a
Hernquist ($\gamma = 1$) model with $r_{a} = 0.3$. The same quantity is given
for simulations with a shorter timestep $\Delta t = 0.01$, and a potential
expansion with $n_{\rm max} = 6$ and $l_{\rm max} = 4$.
\label{fig:additional}}
\end{figure}

We did additional runs to verify that our results were indeed independent of
the timestep and the number of terms used in the potential expansion. Figure
\ref{fig:additional} shows the evolution of the minor to major axis ratio for
the Hernquist ($\gamma = 1$) model with anisotropy radius $r_{a}=0.3$. The run
for this model was repeated using a shorter timestep, $\Delta t = 0.01$, with
essentially the same behavior for the axis ratio. A run with $n_{\rm max} = 6$
and $l_{\rm max} = 4$ converged to a slightly smaller final axis ratio.
Similar results were obtained for other examples.
 
\subsection{Radial stability} 

The sufficient condition $\partial f/\partial E<0$ (Dor\'emus \& Feix 1973;
Dejonghe \& Merritt 1988) can be used to demonstrate the stability against
radial perturbations of the Osipkov-Merritt $\gamma$-models. The lower value
for the anisotropy radius which satisfies the Dor\'emus-Feix's criterion,
$r_{DF}$, is given in Table \ref{table1}. We have studied the stability to
radial perturbation modes of some of those models that fail to satisfy this
criterion, using the same N-body code described in \S~\ref{sec:code}, but
retaining only the $l=0$ terms in the potential expansion. No signs of radial
instability were observed in these runs.

\begin{figure}[t]
\centerline{\psfig{figure=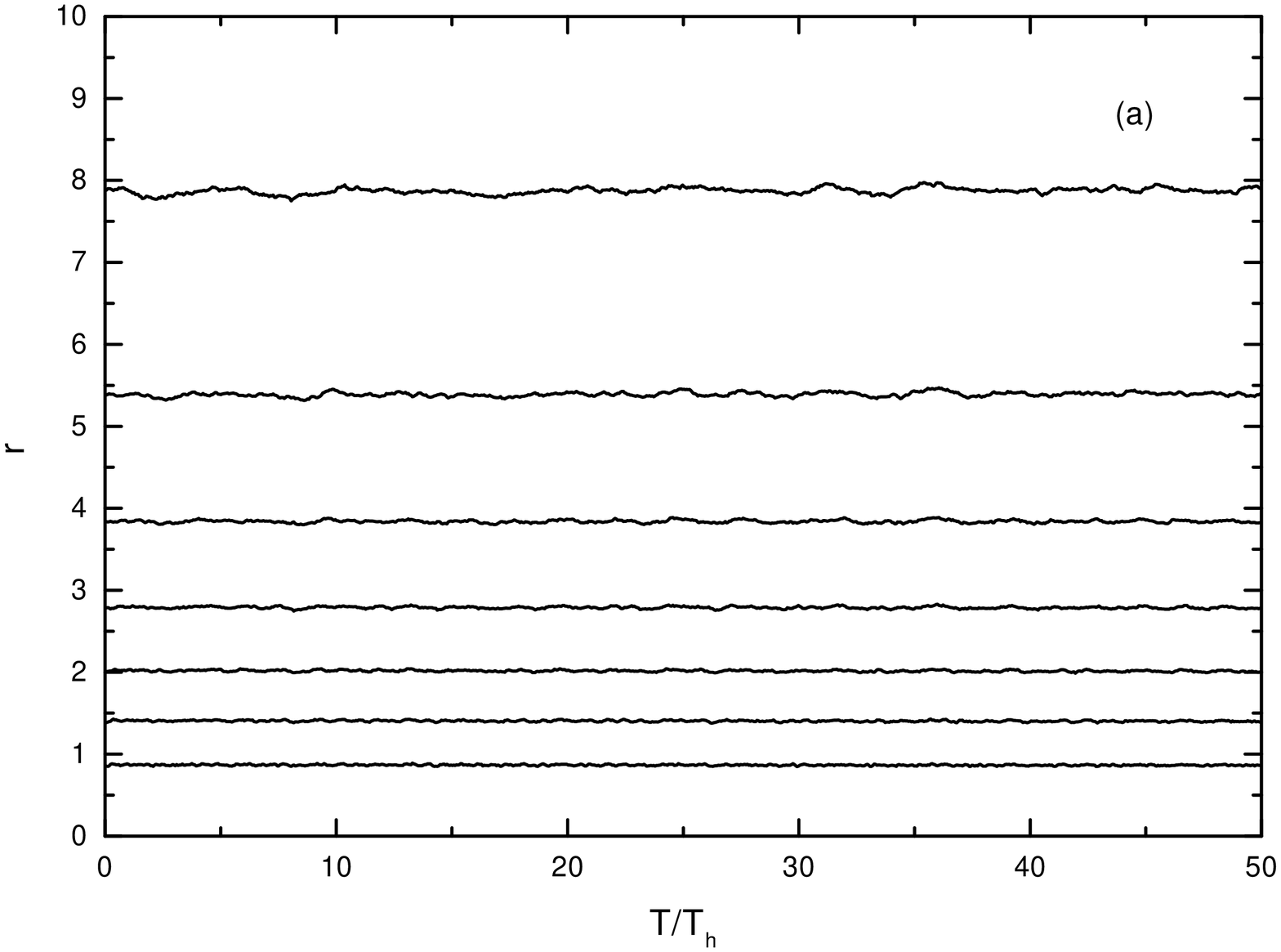,height=0.55\hsize,angle=0}}
\centerline{\psfig{figure=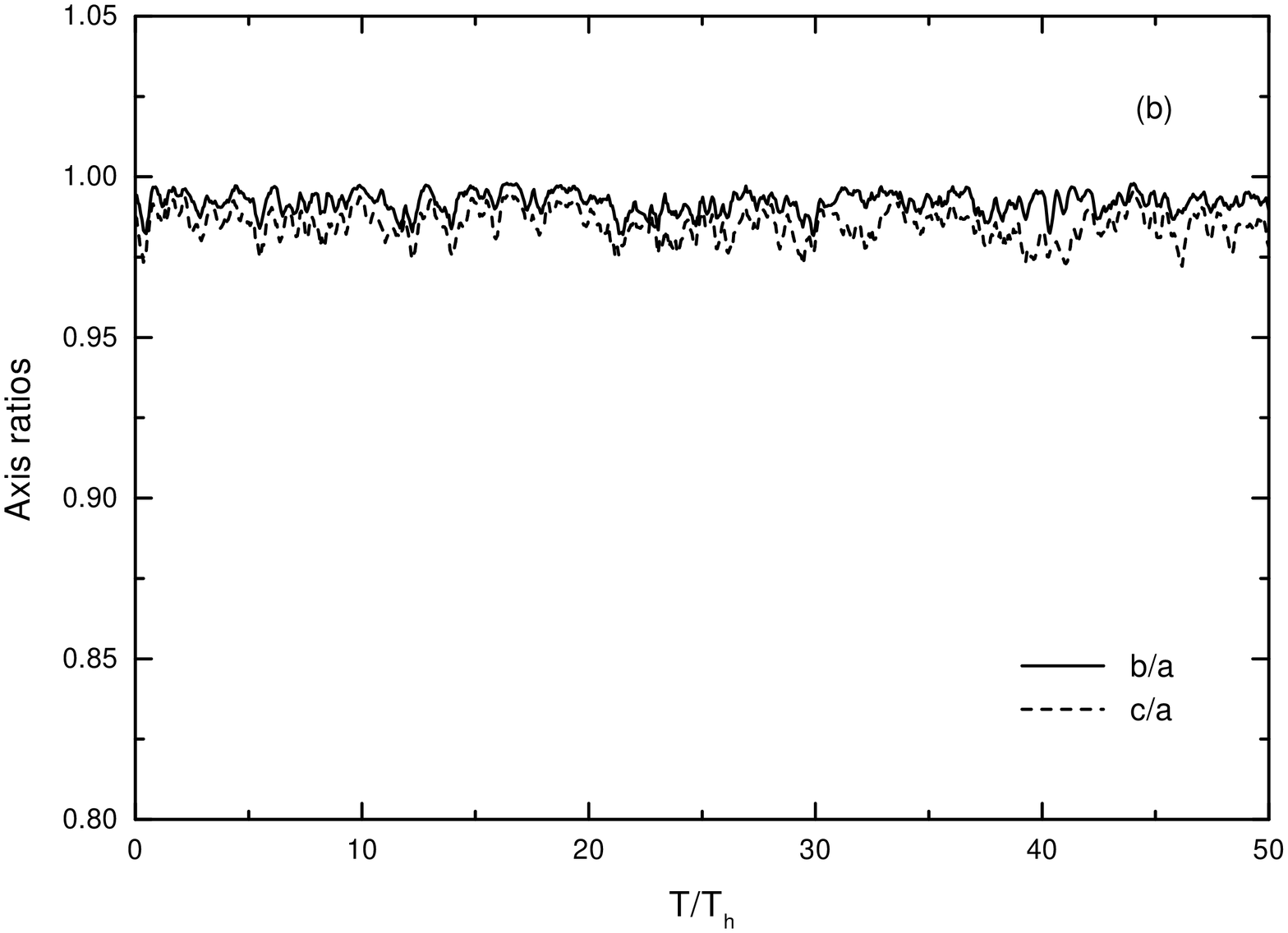,height=0.55\hsize,angle=0}}
\caption{Evolution of the radial distribution of matter (a) and
the axis ratios (b) for the $\gamma = 0$ model with anisotropy radius $r_{a} =
0.5$. These results were obtained retaining only the $l = 0$ terms in the
potential expansion. \label{fig:radial}}
\end{figure}

An example of this result is displayed in Figure \ref{fig:radial}, which is a
plot of the evolution of the radii containing 10\%, 20\%, \ldots, 70\% of the
total mass and the minor to major axis ratio, for the $\gamma=0$ model with
anisotropy radius $r_{a}=0.5$. In both cases, the mass distribution only shows
the fluctuations associated with the finite number of particles. There are no
discernible changes in the radial distribution of matter and the shape of the
particle distribution. Similar results were observed in the other examples,
which do not satisfy the Dor\'emus-Feix's criterion. We conclude that all
these models are stable to radial modes.

\section{Discussion and summary}

\begin{figure}[t]
\plotone{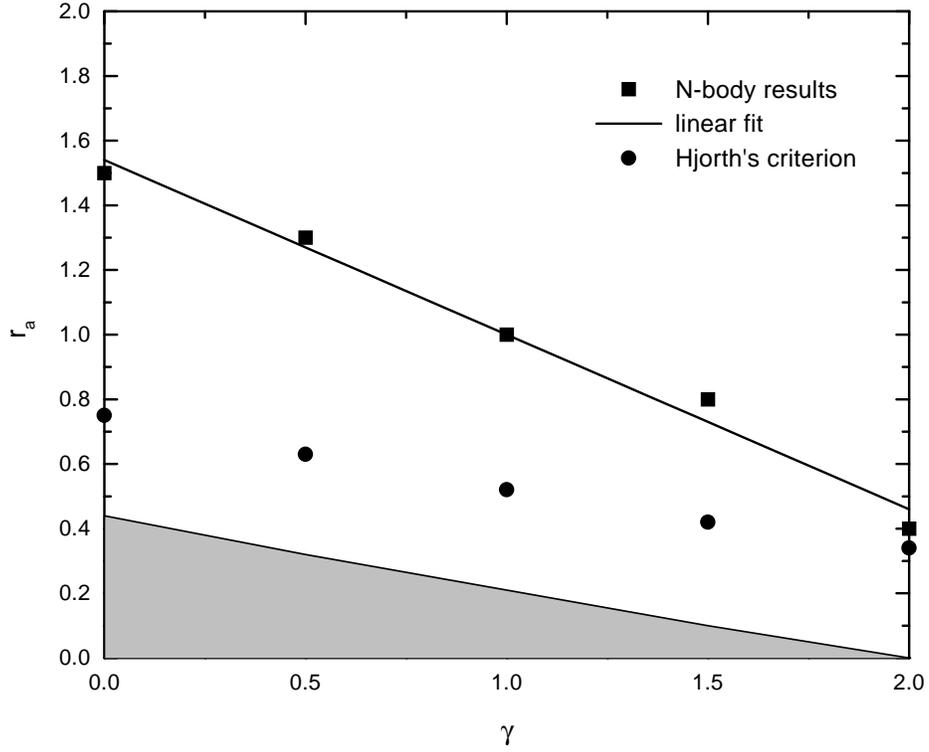}
\caption{The $(\gamma,r_{a})$-plane for Osipkov-Merritt
$\gamma$-models. Models in the shaded region have $f(Q)<0$ for some values of
$Q$. The squares plotted are our numerical results and the
stability threshold predicted by the Hjorth's criterion $df/dQ \le 0$ is
marked with circles. The solid curve drawn is a linear fit obtained with our
numerical results and roughly represents the lower boundary for models which
are not affected by the radial-orbit instability. \label{fig:plane}}
\end{figure}

Recently, Hjorth (1994) has proposed an analytical criterion for the onset
of instability for Osipkov-Merritt models. We have used our simulations to
test this criterion. For the models discussed here, we have found that the
numerical instability thresholds are higher than the predicted value given
by Hjorth's criterion, $df/dQ\le 0$ (circles in Figure \ref{fig:plane}).
Other independent numerical results point in the same direction. Dejonghe \&
Merritt (1988) using a multipolar expansion N-body code have found that the
Osipkov-Merritt-Plummer model is unstable for an anisotropy radius
$r_{a}\lesssim 1.1$, while Hjorth's criterion predicts $r_{a}\lesssim 0.9$.
For this model, we also have tried a different approach doing a series of
simulations using an N-body code based on the Clutton-Brock (1973) basis set
(see also HO92). Our results (Meza \& Zamorano 1996) indicate that the
stability boundary lies near $r_{a} = 1.1$; the same value previously obtained
by Dejonghe \& Merritt (1988).

\begin{figure}[t]
\plotone{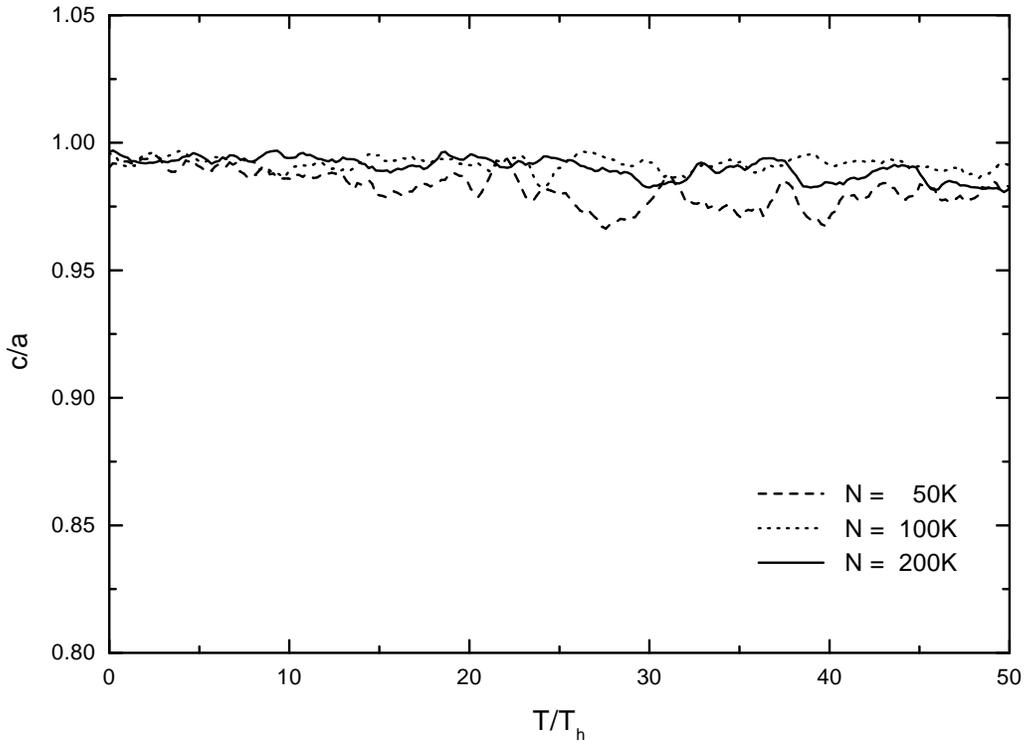}
\caption{Evolution of the minor to major axis ratio for the
$\gamma$ = 3/2 model using $N=5\times 10^4, 10^5$, and $2\times 10^5$
particles and anisotropy radius $r_{a}=0.8$. These values were obtained
fitting an ellipsoid at radius $r_{\rm m}=3.7$. \label{fig:nodep}}
\end{figure}

According to Hjorth (1994), this discrepancy may be originated by numerical
errors in the representation of the central potential due to the small
number of particles used in the simulations. He argues that these errors
could induce artificial inflection points in the DF of the N-body
realization. In this case, there should be a correlation between the number
of particles and the stability threshold determined using these conditions:
fewer particles should start an instability for higher values of the
anisotropy radius $r_{a}$. To decide about this eventual behavior we have
done some additional runs with $N=10^5$ and $2\times 10^5$ particles for
anisotropy radius closer to the critical value quoted for the
$\gamma$-models studied in the previous section. Figure \ref{fig:nodep}
shows the results for the evolution of the minor to major axis ratio for the
$\gamma$ = 3/2 model with anisotropy radius $r_{a}=0.8$. We observe that the
curves show basically the same behavior and we conclude that the number of
particles does not affect our estimation for the stability threshold. The
same result repeats for the other $\gamma$-models. After these
considerations, we feel that it would be interesting to review the
hypothesis used in the Hjorth's criterion.

Our results can be summarized as follows:
\begin{enumerate}
\item Using an N-body code based on the `self-consistent field' method
developed by HO92, we have tested the stability of a set of $\gamma$-models
with DFs of the Osipkov-Merritt type. We found an approximated stability
boundary for the ROI on the $(\gamma,r_{a})$-plane. This boundary is given
by a global anisotropy parameter $2T_{r}/T_{t} = 2.31$.

\item The criterion $df/dQ\le 0$, recently suggested by Hjorth, gives 
a lower stability threshold for the $\gamma$-models.

\item We have studied the stability to radial modes of those $\gamma$-models
which fail to satisfy the Dor\'emus-Feix's stability criterion. No signs of
instability were observed and we conclude that these models are stable to
radial (shell forming) modes. 
\end{enumerate}

\acknowledgements 

We are grateful to David Merritt for a very constructive criticism of an early
version of this manuscript. We thank to the referee, Tim de Zeeuw, for comments
on the manuscript that helped us to improve our discussion in section \S\ 3.
We also wish to thank Jens Hjorth for useful conversations and E-mail 
correspondence. This work was supported by a grant from CONICYT and 
FONDECYT 2950005 (AM) and DTI E-3646-9312 and FONDECYT 1950271 (NZ).

\clearpage

\begin{deluxetable}{ccccccrcc}
\footnotesize
\tablecaption{Parameters of the models\label{table1}}
\tablewidth{13cm}
\tablehead{ 
\colhead{$\gamma$}   & \colhead{$r_{h}$}    & \colhead{$r_{0}$} &
\colhead{$r_{DF}$}   & \colhead{$r_{a}$}    & \colhead{$2T_{r}/T_{t}$} &
\colhead{$T_{h}$}    & \colhead{$\Delta t$} & \colhead{$n_{\rm max}$}
}
\startdata
0   &  3.85 & 0.44 & 0.75 & 1.5 & 2.63 & 16.8 & 0.05 &  4  \nl
1/2 &  3.13 & 0.32 & 0.63 & 1.3 & 2.47 & 12.3 & 0.02 &  6  \nl
1   &  2.41 & 0.20 & 0.52 & 1.0 & 2.39 &  8.3 & 0.02 &  4  \nl
3/2 &  1.70 & 0.10 & 0.42 & 0.8 & 2.09 &  4.9 & 0.01 &  6  \nl
2   &  1    & 0    & 0.34 & 0.4 & 1.98 &  2.2 & 0.01 &  14 \nl
\enddata
\end{deluxetable}

\clearpage 

\clearpage 

\clearpage

\clearpage

\clearpage

\begin{figure}[t]
\caption{Particle positions as viewed along the minor
semiaxis at the beginning (a) and at the end (b) of a simulation for the
Hernquist model with $r_{a}=0.3$. The particle positions are rotated such that
the major (intermediate) semiaxis remains aligned with the $x$-axis
($y$-axis). This region contains approximately 70\% of the total mass of the
system. \label{fig:projection}}
\end{figure}

\clearpage

\clearpage

\end{document}